# Piezoelectric Manipulation and Engineering for Layertronics in Two-Dimensional Materials

Jianke Tian[1], Jia Li*[,1,2], Hengbo Liu[1], Yan Li[1], Ze Liu[1], Linyang Li[1], Jun Li[1], Guodong Liu[1], Junjie Shi[3]

[1]School of Science, Hebei University of Technology, Tianjin 300401, People's Republic of China

[2]College of Science, Civil Aviation University of China, Tianjin 300300, People's Republic of China

[3]State Key Laboratory for Artificial Microstructures and Mesoscopic Physics, School of Physics, Peking University Yangtze Delta Institute of Optoelectronics, Peking University, 5 Yiheyuan Street, Beijing, 100871, People's Republic of China.

**ABSTRACT:**

The electronic transport characteristics of two-dimensional (2D) systems have widespread application prospects in the fabrication of multifunctional nanodevices. However, the current research for basic transport phenomena, such as anomalous valley Hall effect (AVHE) and piezoelectric response, is limited to discrete discussion. Here, we theoretically propose a valley-piezoelectricity coupling strategy beyond the existing paradigm to realize AVHE and layer Hall effect (LHE) in ferrovalley (FV) systems, and its essential principle can be extended to general valleytronic materials. Through first-principles calculations, we demonstrate that the large polarized electric field of $2.8 \times 10^6$ ($1.67 \times 10^7$) V/m can be induced by 0.1% uniaxial strain in FV 2$H$-LaHF (1$T$-LaHF) monolayers. In addition, the microscopic mechanism of interlayer

---

*Corresponding author. *E-mail address*: lijia@cauc.edu.cn (J. L.).

antiferromagnetic (AFM) state of 2*H*-LaHF bilayer is uncovered by the spin Hamiltonian and super-superexchange (SSE) interaction. Our findings pave the way for new explorations of valley Hall-related effect involving piezoelectricity.

**KEYWORDS:** *piezotronics, valley polarization, valley-related muptiple Hall effect, stacking order, super-superexchange*

Berry curvature, which represents the entanglement of the bands, describes the geometric properties of the wave function.[1-3] The Hall effect originates from the coupling of Berry curvature with multiple electronic degrees of freedom, including charge, spin, valley and etc.[4-9] The spin degree of freedom is associated with Berry curvature in time-reversal ($T$) broken system to generate the quantum anomalous Hall effect (QAHE),[2-6] while the space-inversion ($P$) symmetry breaking in valleytronics causes valley Hall effect (VHE) to establish a connection linking unequal Berry curvature with the valley degree of freedom.[10-13] The $T$- and $P$-broken make AVHE and piezoelectric polarization often coexist in a single FV system, and the novel coupling of valley and piezoelectricity, i.e. unconventional piezoelectricity and Berry-Curvature Dopole has aroused widespread concern.[14-16] These findings suggest a rich connection between valley, Berry curvature, and piezoelectricity, however, the transport relationship between piezoelectric response and valley Hall-related effect are still lacking.[10-20]

2D van der Waals (vdW) materials bring novel degree of freedom for electronic engineering due to the weak interlayer vdW interaction, and flexible vdW stacking releases highly tunable magnetic and electronic properties.[20-27] The unique layer degree of freedom and Berry curvature are interrelated to form the LHE, in which electrons in different layers experience opposite deflections due to the influence of the unequal Berry curvature.[28] The sliding ferroelectricity via interlayer sliding in $YI_2$ and $GdI_2$ bilayers,[29,30] as well as the layer-polarized anomalous Hall effect (LP-AHE) caused by stacking orders in $VSi_2P_4$ and $MnBi_2Te_4$ bilayers,[31,32] exhibits rich physical phenomena in layertronics. Therefore, it is very meaningful to explore the transport characteristics of monolayer and bilayer FV systems and the effect of stacking order on the valley behavior and intrinsic magnetism.

In this work, we propose the phenomena of novel piezoelectric anomalous valley Hall effect (PAVHE) and piezoelectric layer Hall effect (PLHE) by coupling the piezoelectric response, valley and layer degrees of freedom, which can be classified as new members of valley-related multiple Hall effect. The coexistence of large spontaneous valley polarization and tunable piezoelectric polarization makes 1$T$-

LaHF and 2*H*-LaHF monolayers be ideal platforms for PAVHE. Then we take 2*H*-LaHF monolayer as an example to study the effect of bilayer stacking orders on magnetism, valley behavior and LHE. Moreover, the microscopic mechanism of interlayer magnetism is revealed by the spin Hamiltonian and interlayer electrons hopping. This work greatly enriches the research on valley Hall-related effect, piezotronics and layertronics.

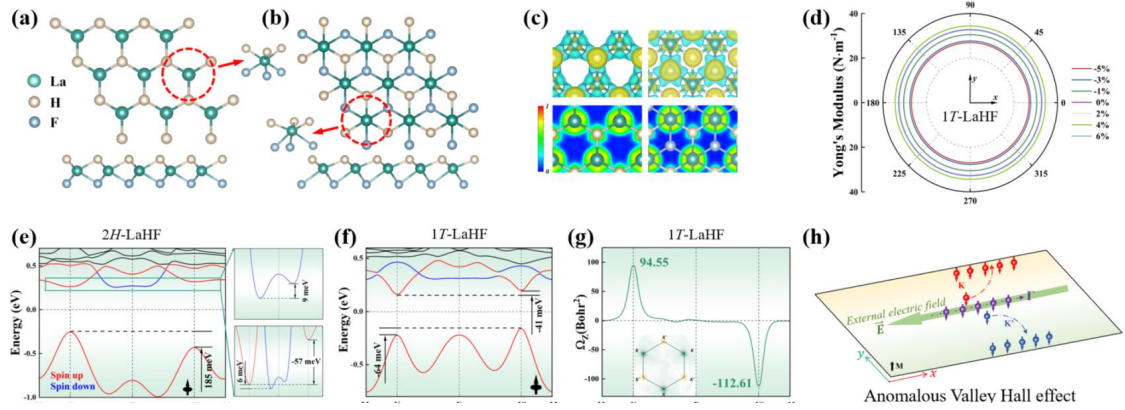

**Figure 1.** Crystal structures of monolayer (a) 2*H*-LaHF and (b) 1*T*-LaHF from top and side views. The unit cell is marked by the red dashed lines. (c) The electron localization function and differential charge densities of 2*H*-LaHF and 1*T*-LaHF. (d) The Young's modulus for 1*T*-LaHF. The electronic band structures of (e) 2*H*-LaHF and (f) 1*T*-LaHF with SOC, and the black arrow indicates the magnetic moment direction of La along the z-axis. (g) The Berry curvatures of 1*T*-LaHF. (h) The diagram of AVHE with K, K' and Γ valleys carriers.

Figure 1a,b exhibits the crystal structures of 2*H*-LaHF and 1*T*-LaHF with *P*-broken and a space group of *P*3*m*1 (No. 156). The in-plane lattice constant for the fully optimized species for 2*H*-LaHF and 1*T*-LaHF are 3.71 and 3.82 Å, respectively. As shown in Figure 1c, the electron localization function shows that the electrons are mainly dispersed around the La, H, and F atoms, indicating their ionic bonding characteristics. Compared with the Young's modulus of typical 2D materials such as graphene (350 N/m) and 2*H*-BN (270 N/m), 2*H*-LaHF and 1*T*-LaHF monolayers exhibit lower stiffness and greater flexibility (in Figure 1d,S1). The excellent stabilities of 2*H*-LaHF and 1*T*-LaHF monolayers was demonstrated by the results of phonon spectrum, molecular dynamic calculation and elastic constants, respectively, as shown in Figure S2. The energy difference between AFM and ferromagnetic (FM)

ordering (in Figure S3) indicates the FM coupling is dominant, which can be explained by Goodenough-Kanamori-Anderson rules.[33] In Figure S4, 2*H*-LaHF (1*T*-LaHF) exhibits the XY ferromagnets property, and the out-of-plane magnetization, which is conducive to generating spontaneous valley polarization, can be realized by overcoming the energy barrier of 787 (33) μmV for 2*H*-LaHF (1*T*-LaHF).

With Spin-orbit-coupling (SOC) effect (in Figure 1e,f and S5), based on the spin channel at the valence band maximum (VBM) and conduction band minimum (CBM), the 2H-LaHF and 1T-LaHF can be regarded as bipolar-semiconductor and half-semiconductor, respectively. It is worth noting that the large spin bandgap of 4.39 and 0.37 eV can be found in 1*T*-LaHF, which can generate 100% spin polarized carriers by electrical gating. Moreover, the valley splitting of 2*H*-LaHF and 1*T*-LaHF at CBM (VBM) is -57 (185) and -41 (-64) meV, respectively, which is larger than or comparable to those in VSe$_2$ (78.2 meV),[34] LaBr$_2$ (33 meV)[35] and MoTe$_2$/EuO (20 meV).[36] The effects of biaxial strain on valley splitting, band gap, MAE (magnetic anisotropic energy) and magnetic moment are shown in Figure S6. The *P*-broken results in unequal nonzero Berry curvature at polarized-valleys (K and K' valleys), and the Berry curvature can be obtained from the Kubo formula:[37,38] $\Omega(k) = -\sum_n \sum_{n \neq n'} f_n \frac{2Im\langle\psi_{nk}|\hat{v}_x|\psi_{n'k}\rangle\langle\psi_{n'k}|\hat{v}_y|\psi_{nk}\rangle}{(E_n - E_{n'})^2}$ where $f_n$ represents the Fermi-Dirac distribution. $|\psi_{nk}\rangle$ and $\hat{v}_x$ ($\hat{v}_y$) are the Bloch wave function with eigenvalue $E_n$ and the velocity operator along $x$ ($y$) direction, respectively. As shown in Figure 1g and S7, the opposite signs and unequal magnitudes of polarized valleys show valley contrasting characteristics of systems. The in-plane external electric field drives the carrier to obtain the group velocity $v_\parallel$, while the Berry curvature give carrier the transverse anomalous velocity $v_\perp$, thus producing the AVHE (in Figure 1h).

In FV system, the *P*-broken results in the piezoelectric polarization as well as the unequal Berry curvature and valley polarization. A natural question is whether there is a coupled transport relationship between valley degree of freedom and piezoelectric response. The orthorhombic supercell be used to calculate the piezoelectric tensor $e_{ik}$ of systems (see Figure S8 a,b). As shown in Figure 2b, the electronic and ionic contributions for piezoelectric tensor $e_{ik}$ have opposite signs and similar values under non biaxial strain, which directly leads to the weak in-plane and out-of-plane piezoelectric response. In Figure. S8 d,e, with the application of tensile strain, the ionic contribution decreases rapidly while the electronic contribution is robust for piezoelectric tensor $e_{11}$. As shown in Figure 2c, the in-plane piezoelectric coefficients $d_{11}$ of 5.41 (10.03) pm/V can be found in 2$H$-LaHF (1$T$-LaHF) under +6% biaxial strain, which is greater than most 2D Janus materials (WSTe (3.33 pm/V), SnSSe (2.25 pm/V) and $MoS_2$ (3.73 pm/V)).[39-41] The $d_{11}$ of 2$H$-LaHF (1$T$-LaHF) without biaxial strain is -0.46 (4.45) pm/V, which indicates that the direction of piezoelectric polarization is opposite (same) to the direction of uniaxial strain ($\vec{\eta}$).

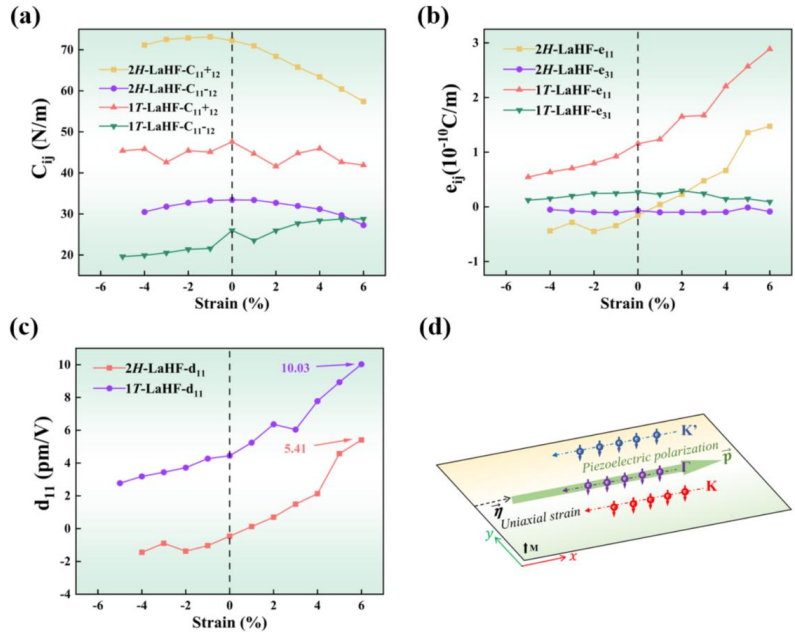

**Figure 2.** (a) The elastic constants $C_{jk}$, (b) piezoelectric tensor $e_{ij}$ and (c) piezoelectric strain coefficient $d_{ij}$ as a function with biaxial strain. (d) The diagram of piezoelectric response induced by $\vec{\eta}$.

The use of hydrostatic pressure and nonconducting atomic force microscopy makes it possible to apply local stress to the sample. Figure 2d exhibits the strain-induced piezoelectric polarization, and the related mechanism is that the polarized electric field induced by $\vec{\eta}$ excites carriers from VBM to CBM to produce the electronic output. Next, we use charge droping, which is widely used as a regulatory means, to selectively excite the carriers of the polarized-valleys. Next, we focus on the intrinsic polarized electric field induced by $\vec{\eta}$ in piezoelectric response. The permittivity, Young's modulus and in-plane piezoelectric coefficients of 2$H$-LaHF (1$T$-LaHF) are 3.53 (4.96), 45.68 (32.68) N/m and -0.46 (4.45) pm/V, respectively. When +0.1% $\vec{\eta}$ is applied, the corresponding stress $\sigma$ can be expressed as $\sigma = Y \cdot \varepsilon$, where $Y$ and $\varepsilon$ are Young's modulus and strain, respectively. The intrinsic polarized electric field $E$ can be expressed as $E = \frac{d_{11} \cdot \sigma}{\varepsilon_{xx} \cdot \varepsilon_0}$, where $\varepsilon_0$ is permittivity of vacuum.

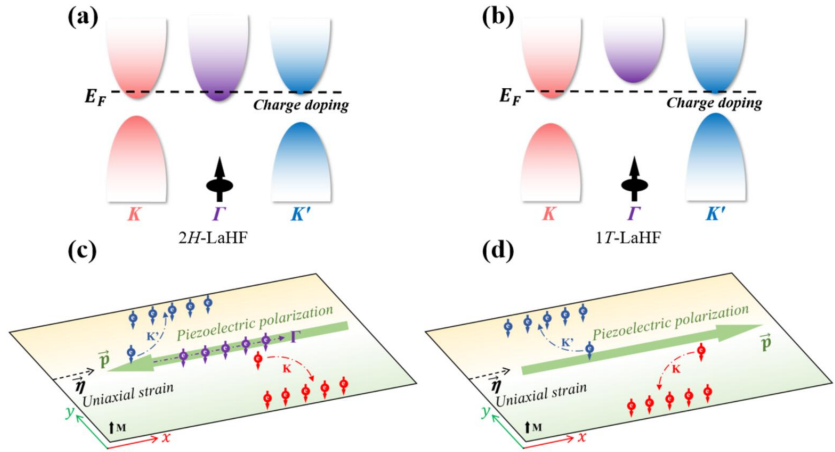

**Figure 3.** The illustration of band structure of PAVHE with Fermi energy levels penetrating polarized-valleys and Γ valley in (a) 2$H$-LaHF and polarized-valleys in (b) 1$T$-LaHF. The diagram of PAVHE in (c) 2$H$-LaHF and (d) 1$T$-LaHF. The direction of stress is opposite (same) to the direction of polarization in 2$H$-LaHF (1$T$-LaHF). The carriers from K, K' and Γ valleys are indicated using red, blue and purple markers, respectively, with arrows marking the spin orientations.

Based on the above formula, 0.1% $\vec{\eta}$ can generate an intrinsic polarized electric field of 2.8×10$^6$ (1.67×10$^7$) V/m in 2$H$-LaHF (1$T$-LaHF), which indicates the high efficiency of $\vec{\eta}$ on the induction of polarization electric field. As shown in Figure 3c,d, carrier doping can effectively regulate Fermi levels to selectively excite polarized-valleys carriers. The polarized electric field and the unequal Berry curvature cause the polarized valley carriers to obtain $v_\parallel$ and $v_\perp$, respectively. The transport phenomenon in which the intrinsic polarized electric field replaces the external electric field to drive the carrier is called PAVHE, which can be classified as one of the members of valley-related multiple Hall effect. Based on the electronic band structure, the PAVHE in 2$H$-LaHF and 1$T$-LaHF under different biaxial strain is shown in Figure S9.

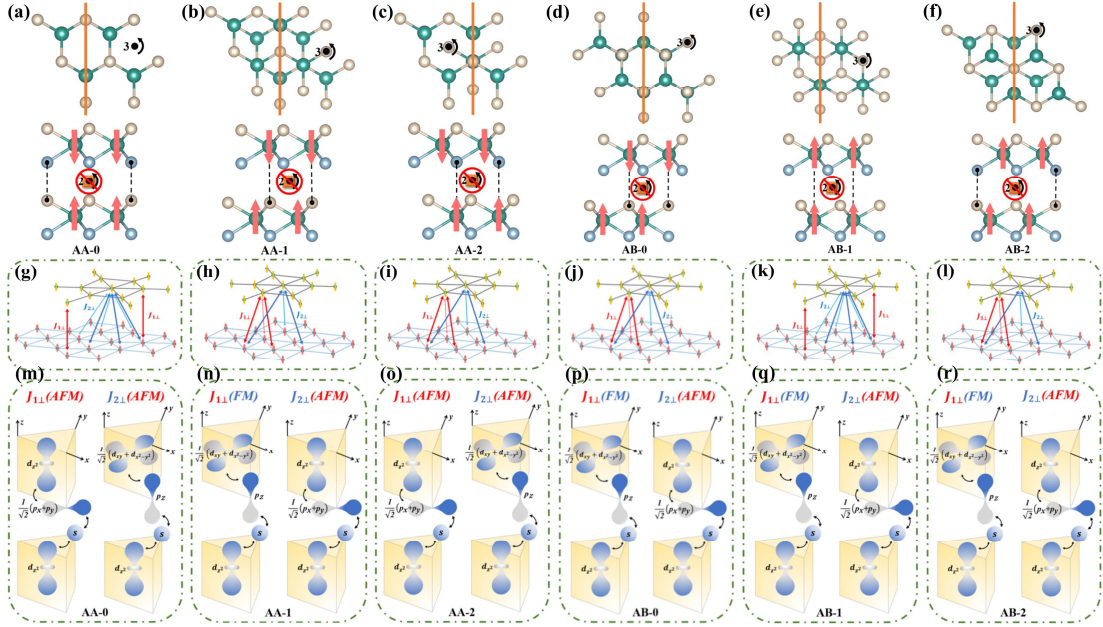

**Figure 4.** (a)-(f) The stacking structures for different stacking orders, and the red arrows represent the direction of magnetization of the La atoms. (g)-(l) The interlayer La nearest neighbor and second-nearest neighbor correspond to $J_{1\perp}$ (red) and $J_{2\perp}$ (blue) for different stacking orders. (m)-(n) Schematics of SSE for different stacking orders.

The stacking orders of bilayer magnetic materials may cause intriguing valley behavior and LHE. We discuss the novel physical behaviors by taking the different stacking sequences (AA-0, AA-1, AA-2, AB-0, AB-1 and AB-2) under the typical stacking of 2$H$-LaHF as an example (in Figure 4a-f). The AA-0 stacking of 2$H$-LaHF bilayer is constructed by placing one layer facing the other layer, resulting in a mirror symmetry and 3-fold rotation axis in $xy$ plane. AA-1 and AA-2 stackings are constructed by interlayer translation operations of $t_{1\parallel}\left[\frac{2}{3},\frac{1}{3},0\right]$ and $t_{1\parallel}\left[\frac{1}{3},\frac{2}{3},0\right]$, respectively. AB-0 stacking can be constructed by rotating the upper layer 180° based on AA-0 stacking, while AB-1 and AB-2 stacking can be obtained from AB-0 stacking by translation operations of $t_{1\parallel}\left[\frac{2}{3},\frac{1}{3},0\right]$ and $t_{1\parallel}\left[\frac{1}{3},\frac{2}{3},0\right]$, respectively. In short, the bilayer systems conventionally constructed by Janus 2$H$-LaHF leads to the broken mirror symmetry along the $z$ direction, which is the basis for the emergence of valley polarization and LHE.[25-27]

As shown in Figure S10, due to the breaking of mirror symmetry, the asymmetrical redistribution of charge density of layer-1 and layer-2 indicates the existence of out-of-plane electrical polarization. Additionally, all stacking of 2$H$-LaHF exhibit the interlayer AFM coupling (M↑↓), while AA-2 and AB-0 stacking exhibit the lowest energy, as indicated in Figure S11a. To understand the interlayer AFM coupling caused by stacking, a simple spin Hamiltonian is established:[42]

$$H = E_0 + \sum_{i,j} J_\parallel S_i \cdot S_j + \sum_{i,k} J_{1\perp} S_i \cdot S_k + \sum_{i,j} J_{2\perp} S_i \cdot S_l, \qquad (1)$$

where $E_0$ is the ground state energy, and $S_i$, $S_j$, $S_k$ and $S_l$ are the magnetic moments at sites $i$, $j$, $k$ and $l$, respectively. In Figure 4g-l, $J$ is the Heisenberg exchange parameter, and $J_\parallel$, $J_{1\perp}$ and $J_{2\perp}$ are the intralayer, interlayer nearest-neighbor (NN) and interlayer next-nearest-neighbor (next-NN) La-La exchange interactions, respectively. The spin Hamiltonian of the relevant stacking and the calculation results of the relevant parameters are presented in Supporting Information and Table S2. $J_\parallel < 0$ indicates that the intralayer exchange is FM coupling, while the signs of $J_{1\perp}$ and $J_{2\perp}$ maybe vary with the stacking types, which reflects the change of the competition relationship between interlayer exchange interaction $J_{1\perp}$ and $J_{2\perp}$.

In the trigonal-prismatic crystal field, the La-$d$ orbitals are split into $d_{z^2}$, $d_{xz}/d_{yz}$ and $d_{xy}/d_{x^2-y^2}$ orbitals. Figure S11b shows the different interlayer exchange interactions of La atoms. The $d_{z^2}$-$d_{z^2}$ hopping is prohibited in FM alignment (blue), whereas it is allowed in AFM alignment (red), suggesting that the $d_{z^2}$-$d_{z^2}$ hopping causes the predominantly AFM coupling.[43] Based on the local Hund coupling, the $d_{z^2}$-$d_{xz}/d_{yz}$ and $d_{z^2}$-$d_{xy}/d_{x^2-y^2}$ hopping mainly contribute to the FM coupling.[33,44] The La-La interlayer exchange interactions are the SSE interaction mediated by H-$s$ and F-$p$ orbitals.

In the following, we quantitatively analyze the competitive relationship between $J_{1\perp}$ and $J_{2\perp}$ under different stackings. The interlayer La-La NN exchange interaction $J_{1\perp}$ (red) and La-La next-NN exchange interaction $J_{2\perp}$ (blue) for AA-1 stacking is shown in Figure 4h. Using H-*s* and F-*p* orbitals as the mediums, Figure 4n shows the $J_{1\perp}$ is associated with the virtual excitations between half-filled La-$d_{z^2}$ and empty $d_{xy}/d_{x^2-y^2}$ orbitals, which results in the interlayer FM coupling. However, $J_{2\perp}$ predominates the interlayer AFM coupling through virtual excitations between the half-filled La-$d_{z^2}$ of layer-1 and layer-2. The AA-1 stacking contains three $J_{1\perp}$ bonds of -0.052 meV and three $J_{2\perp}$ bonds of 0.804 meV per unit cell, respectively, which means that next-NN interlayer exchange interactions dominates the AFM state for AA-1 stacking. For AB-1 stacking, the complete coincidence of the relative positions between layer-1 and layer-2 causes the number of $J_{1\perp}$ ($J_{2\perp}$) bonds become one (six). Although the absolute value of $J_{1\perp}$ bonds (-0.475 meV) is greater than $J_{2\perp}$ bonds (0.115 meV), the number of $J_{2\perp}$ bonds is much greater than $J_{1\perp}$ bonds, which leads to the emergence of the interlayer AFM state. Although both AA-2 and AB-2 have three $J_{1\perp}$ and $J_{2\perp}$ bonds, their energy differences of FM and AFM states are quite different (see Table S2), which should be attributed to the symmetry difference (in Figure 4c,f).[30]

In the following, the electric band structures of ground-state stacking configurations (AA-2 and AB-0) are investigated based on the AFM magnetic ground state. The valley polarization can be found at VBM at K and K' points and CBM at K-Γ and K'-Γ paths (in Figure 5a,b). Under the influence of intrinsic ferromagnetic sequence and SOC effect, VBM carriers comes from the spin-down electrons of layer-2, while CBM carriers are mainly contributed by the spin-down electrons of layer-1, indicating that AA-2 and AB-0 stacking are indirect bandgap half-semiconductors. The valley polarizations of AA-2 and AB-0 stacking at VBM (CBM) are -168 (-19) and 174 (-13) meV under M↑↓ magnetic configuration, respectively. As shown in Figure S12, unequal Berry curvature of AA-2 (AB-0) stacking at K and K' valleys are 16.01 (6.72) and -18.47 (-9.06) Bohr$^2$, respectively.

Considering the contribution of different layers to the polarized-valleys, different interlayer carriers can be selectively excited by carrier doping. As shown in Figure 5c,d, appropriate hole doping can excite K' valley carriers of layer-2, while the K-Γ path carriers of layer-1 can be induced by charge doping. The direction of the intrinsic polarized electric field induced by piezoelectric polarization in 2*H*-LaHF without biaxial strain is opposite to the direction of $\vec{\eta}$. In Figure 5c, when the inward $\vec{\eta}$ is applied to layer-2, the hole carriers move outward driven by the polarized electric field and converge to the left side of the sample under the action of Berry curvature. Charge doping excites the CBM1 valley at K-Γ path, and berry curvature and outward uniaxial strain cause charge carriers to converge on the right side of the sample. As shown in Figure 5 and S13, with interlayer sliding, the valley polarization of VBM mostly exceeds 160 meV, while the valley of CBM transitions between K (K') and K-Γ path, suggesting 2*H*-LaHF bilayer potential applications in layertronics.

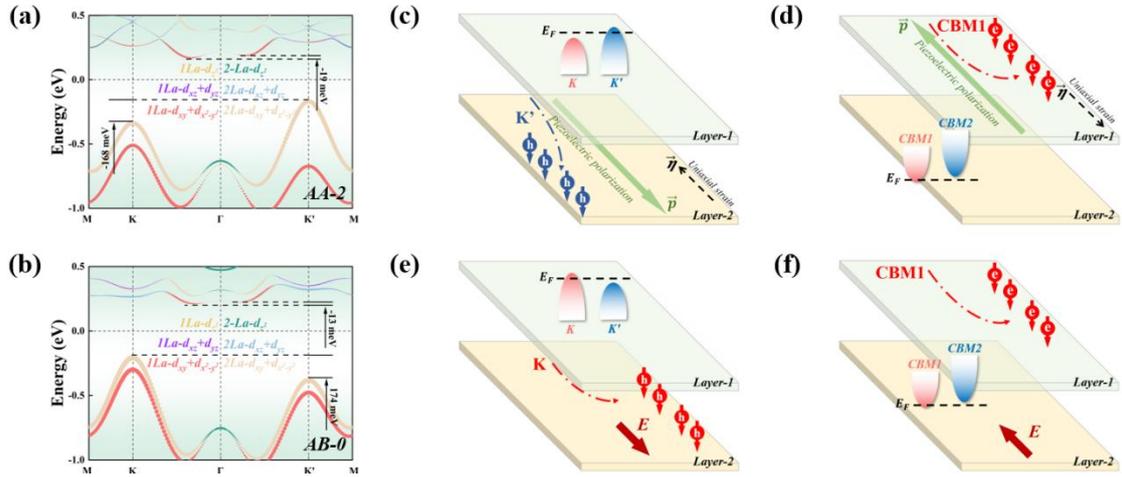

**Figure 5.** The electric band structures of (a) AA-2 and (b) AB-0. Diagrams of transport phenomenon under (c) hole and (d) charge doping for AA-2 stacking. Diagrams of transport phenomenon under (e) hole and (f) charge doping for AB-0 stacking. (c), (d) and (e), (f) shows the LHE induced by the piezoelectric response and in-plane electric field, respectively. The carriers from K and K' valleys of different layers are indicated using red and blue markers, respectively, with arrows marking the spin orientations.

Similar to the relationship between PAVHE and AVHE, we theoretically propose the concept of PLHE involving piezoelectric response on the basis of LHE, which also can be classified as a new member of valley-related multiple Hall effect. There is no layer-related phenomenon such as sliding ferroelectricity and LP-AHE in 2*H*-LaHF bilayer,[29-32] however, the existence of piezoelectric polarization provides a new means of regulating Bloch electrons in PAVHE and PLHE. Although we primarily focused on trigonal lattices (*P*3*m*1) to discuss PAVHE and PLHE, the principles of essentiality can also be adapted to other lattice structures, such as square lattice GaBr and GeSe monolayers (*P*4/*nmm*) and altermagnetic material CrO (*P*4/*mmm* space group).[45-47]

In conclusion, we theoretically propose a novel mechanism to realize piezoelectric AVHE and LHE based on FV 2*H*-LaHF and 1*T*-LaHF monolayers. The coexistence of unequal Berry curvature and piezoelectric response under *P*-broken makes the valley-piezoelectricity coupling mechanism is universal in 2D valleytronic materials. Based on first-principles calculations, the large intrinsic polarized electric field of $2.8 \times 10^6$ ($1.67 \times 10^7$) V/m can be induced by 0.1% uniaxial strain in 2*H*-LaHF (1*T*-LaHF). Furthermore, the microscopic mechanism of interlayer AFM state in 2*H*-LaHF bilayer is revealed by spin Hamiltonian and interlayer electron hopping. Our work has great significance for the fundamental research of valley Hall-related effect, and may facilitate experimental work related to valleytronics and piezophotonics.

## ASSOCIATED CONTENT

**Supporting Information**

Details of the calculation methods, spin Hamiltonian model, Heisenberg exchange parameters, phonon dispersion spectrum, MAE and Curie temperature, electronic and piezoelectric property, PAVHE and PLHE, the energy difference between interlayer AFM and FM states for 2*H*-LaHF bilayer.

## AUTHOR INFORMATION

**Corresponding Authors**

Jia Li - *School of Science, Hebei University of Technology, Tianjin 300401, People's Republic of China; College of Science, Civil Aviation University of China, Tianjin 300300, People's Republic of China;* E-mail: lijia@cauc.edu.cn

**Authors**

**Jianke Tian -** *School of Science, Hebei University of Technology, Tianjin 300401, People's Republic of China*

**Hengbo Liu -** *School of Science, Hebei University of Technology, Tianjin 300401, People's Republic of China*

**Yan Li -** *School of Science, Hebei University of Technology, Tianjin 300401, People's Republic of China*

**Ze Liu -** *School of Science, Hebei University of Technology, Tianjin 300401, People's Republic of China*

**Linyang Li -** *School of Science, Hebei University of Technology, Tianjin 300401, People's Republic of China*

**Jun Li -** *School of Science, Hebei University of Technology, Tianjin 300401, People's Republic of China*

**Guodong Liu -** *School of Science, Hebei University of Technology, Tianjin 300401, People's Republic of China*

**Junjie Shi -** *State Key Laboratory for Artificial Microstructures and Mesoscopic Physics, School of Physics, Peking University Yangtze Delta Institute of Optoelectronics, Peking University, 5 Yiheyuan Street, Beijing, 100871, People's Republic of China.*

**Notes**

The authors have no conflicts of interest to disclose.

**ACKNOWLEDGMENTS**

This work was supported by the Natural Science Foundation of Hebei Province (No. A2020202010).

# REFERENCES

(1) Wimmer, M.; Price, H.; Carusotto, I.; Peschel, U. Experimental Measurement of the Berry Curvature from Anomalous Transport. *Nat. Phys.* **2017,** 13, 545.

(2) Xiao, J.; Wang, Y.; Wang, H.; Pemmaraju, C. D.; Wang, S.; Muscher, P.; Sie, E. J.; Nyby, C. M.; Devereaux, T. P.; Qian, X.; Zhang, X.; Lindenberg, A. M. Berry Curvature Memory through Electrically Driven Stacking Transitions. *Nat. Phys.* **2020,** 16, 1028.

(3) Nagaosa, N.; Sinova, J.; Onoda, S.; MacDonald, A.; Ong, N. Anomalous Hall effect. *Rev. Mod. Phys.* **2010,** 82, 1539.

(4) Cohen, E.; Larocque, H.; Bouchard, F.; Nejadsattari, F.; Gefen, Y.; Karimi, E. Geometric phase from Aharonov-Bohm to Pancharatnam-Berry and beyond. *Nat. Rev. Phys.* **2019,** 1, 437.

(5) Deng, Y.; Yu, Y.; Shi, M. Z.; Guo, Z.; Xu, Z.; Wang, J.; Chen, X. H.; Zhang, Y. Quantum anomalous Hall effect in intrinsic magnetic topological insulator $MnBi_2Te_4$. *Science* **2020,** 367, 895.

(6) Liu, Z.; Zhao, G.; Liu, B.; Wang, Z. F.; Yang, J.; Liu, F. Intrinsic Quantum Anomalous Hall Effect with In-Plane Magnetization: Searching Rule and Material Prediction. *Phys. Rev. Lett.* **2018,** 121, 246401.

(7) Mak, K.; McGill, K.; Park, J.; McEuen, P. Valleytronics. The valley Hall effect in $MoS_2$ transistors. *Science* **2014,** 344, 1489.

(8) Hu, H.; Tong, W.-Y.; Shen, Y.-H.; Wan, X.; Duan, C.-G. Concepts of the half-valley-metal and quantum anomalous valley Hall effect. *npj Comput. Mater.* **2020,** 6,129.

(9) Xiao, D.; Liu, G.-B.; Feng, W.; Xu, X.; Yao, W. Coupled Spin and Valley Physics in Monolayers of $MoS_2$ and Other Group-VI Dichalcogenides. *Phys. Rev. Lett.* **2012,** 108, 196802.

(10) Ma, Y.; Kou, L.; Du, A.; Huang, B.; Dai, Y.; Heine, T. Conduction-band valley spin splitting in single-layer H-Tl2O. *Phys. Rev. B* **2018,** 97, 035444.

(11) Zhang, T.; Xu, X.; Huang, B.; Dai, Y.; Ma, Y. 2D spontaneous valley polarization from inversion symmetric single-layer lattices. *npj Comput. Mater.* **2022,** 8, 64.

(12) Zhou, J.; Sun, Q.; Jena, P. Valley-Polarized Quantum Anomalous Hall Effect in Ferrimagnetic Honeycomb Lattices. *Phys. Rev. Lett.* **2017,** 119, 046403.

(13) Ning, Z.; Ding, X.; Xu, D.-H.; Wang, R. Erratum: Robustness of half-integer quantized Hall conductivity against disorder in an anisotropic Dirac semimetal with parity anomaly. *Phys. Rev. B* **2023,** 108, 159901.

(14) Rouzhahong, Y.; Liang, C.; He, J.; Lin, X.; Wang, B.; Li, H. Unconventional Piezoelectricity of Two-Dimensional Materials Driven by the Hall Effect. *Nano Lett.* **2024,** 24, 1137.

(15) Rouzhahong, Y.; Liang, C.; Li, C.; Wang, B.; Li, H. Valley piezoelectricity promoted by spin-orbit coupling in quantum materials. *Sci. China: Phys., Mech. Astron.* **2023,** 66, 247711.

(16) Xiao, R.-C.; Shao, D.-F.; Zhang, Z.-Q.; Jiang, H. Two-Dimensional Metals for Piezoelectriclike Devices Based on Berry-Curvature Dipole. *Phys. Rev. Appl.* **2020,** 13, 044014.

(17) Peng, R.; Ma, Y.; Xu, X.; He, Z.; Huang, B.; Dai, Y. Intrinsic anomalous valley Hall effect in single-layer $Nb_3I_8$. *Phys. Rev. B* **2020,** 102, 035142.

(18) Zhang, H.; Yang, W.; Ning, Y.; Xu, X. Abundant valley-polarized states in two-dimensional ferromagnetic van der Waals heterostructures. *Phys. Rev. B* **2020,** 101, 205404.

(19) Cheng, H.-X.; Zhou, J.; Ji, W.; Zhang, Y.-N.; Feng, Y.-P. Two-dimensional intrinsic ferrovalley $GdI_2$ with large valley polarization. *Phys. Rev. B* **2021,** 103, 125121.